# Time-Varying Metasurfaces and Lorentz Non-Reciprocity

Amr Shaltout, Alexander Kildishev and Vladimir Shalaev[*]

*Birck Nanotechnology Center, School of Electrical and Computer Engineering,
Purdue University, West Lafayette, IN 47907, USA*
[*]shalaev@purdue.edu

**A cornerstone equation of optics – Snell's law – relates the angles of incidence and refraction for light passing through an interface between two media. It is built on two fundamental constrains: the conservation of tangential momentum and the conservation of energy. By relaxing the classical Snell's law photon momentum conservation constrain when using space-gradient phase discontinuity, optical metasurfaces enabled an entirely new class of ultrathin optical devices. Here, we show that by eradicating the photon energy conservation constrain when introducing time-gradient phase discontinuity, we can further empower the area of flat photonics and obtain a new genus of optical devices. With this approach, classical Snell's relations are developed into a more universal form not limited by Lorentz reciprocity, hence, meeting all the requirements for building magnetic-free optical isolators. Furthermore, photons experience inelastic interaction with time-gradient metasurfaces, which modifies photonic energy eigenstates and results in a Doppler-like wavelength shift. Consequently, metasurfaces with both space- and time-gradients can have a strong impact on a plethora of photonic applications and provide versatile control over the physical properties of light.**

Optical metasurfaces have produced a strong impact on nanophotonic devices resulting from the generalization of Snell's relation[1], which has always played a primary role in the design of optical devices. Snell's law relates the angles of incidence and refraction for light passing through an interface between two distinct isotropic media and it is a direct result of two fundamental laws[2] for reflected and refracted photons dealing with the conservation of (i) tangential momentum, and (ii) energy. With the inception of space-gradient metasurfaces, created using nano-antennas that introduce spatially dependent phase-discontinuities along the interface, Snell's law has been generalized to include a discontinuity in the tangential momentum. Consequently, it has already become possible to engineer angles of reflection and refraction at will with flat, ultra-thin metasurfaces. That has enabled numerous applications in a whole new family of flat photonic devices[3] including light bending[1, 4], flat lenses[5-9], ultra-thin holograms[10-15], polarization plates[16, 17] and other exciting physical phenomena including the photonic spin Hall effect[18-20], vortex beam generation[1, 21], bi-anisotropic devices[22, 23], and chiral effects[24-28].

By introducing a dynamic (i.e. temporal) change to the phase-discontinuity, Snell's law can be modified to an even more universal form, which is expected to lead to other new and exciting physical impacts. As we will show, photons interacting with time-gradient metasurfaces do not conserve their energy. As a result, there is a change in the normal momentum component induced by the energy-momentum dispersion relation in the media on both sides of the metasurface. Therefore, using both space and time variation of phase induced by metasurfaces, it is possible to control both the tangential and normal momentum components. Beside the extra degree of freedom of controlling the normal momentum, the modification to Snell's law contributed by a time variation of the phase-shift breaks Lorentz reciprocity[29]. This enables the implementation of optical devices, such as optical isolators, which require breaking time-reversal reciprocity[30, 31] and thus are unachievable with space-gradient metasurface alone.

First, we derive the mathematical apparatus for the generalized reflection and refraction using space and time phase-shift variation along metasurfaces. Then, the non-reciprocity of a new form of Snell's law is highlighted, and possible schematics for designing optical isolators using time-varying metasurfaces are proposed. Finally, new physical effects induced by the time variation of phase discontinuity are discussed and possible implementation techniques are proposed.

Snell's law is a geometric optics approximation, which is exact when we are dealing with ideal plane waves. It still is very accurate, however, if the wave amplitude is slowly varying in space with respect to wavelength scale, and in time with respect to the period. In this case, electromagnetic waves can be viewed as a collection of local rays in space and time. The wave approximation corresponding to geometric optics is given by[29]:

$$\mathbf{E} = \mathbf{a} e^{i\psi} \tag{1}$$

where $\mathbf{a}$ is a slowly varying function of space and time (constant in case of plane waves), and the phase term $\psi$ (also called the eikonal in geometric optics) is a fast varying function of space and time. The spatial and temporal derivatives of $\psi$ give the frequency and wave vector of the wave, respectively[29]:

$$\omega = -\frac{\partial \psi}{\partial t} \tag{2}$$

$$\mathbf{k} = \nabla \psi \tag{3}$$

We study the most general case shown in figure 1, when a wave with a phase of $\psi_i$ is incident on a metasurface, which induces a space-time varying phase-shift of $\psi_{ms,r}$ for a reflected wave and $\psi_{ms,t}$ for refracted (transmitted) wave. This means that the phases of the reflected and transmitted waves are given by:

$$\psi_s = \psi_i + \psi_{ms,s}, \ \ s = \{r,t\}. \tag{4}$$

By applying equations (2) and (3) to both sides of equations (4), we obtain:

$$\omega_s = \omega_i - \partial \psi_{ms,s}/\partial t, \ \ s = \{r,t\}; \tag{5}$$

$$k_{s,x} = k_{i,x} + \partial \psi_{ms,s}/\partial x, \ \ s = \{r,t\}, \tag{6}$$

where $\omega_i$, $\omega_r$, $\omega_t$, $k_{i,x}$, $k_{r,x}$ and $k_{t,x}$ are the frequencies and the x-components of the wave-numbers of incident, reflected and transmitted waves, respectively. Equation (6) can be rewritten in terms of the wavenumbers' amplitudes $k_i$, $k_r$, and $k_t$ as follows:

$$k_r \sin\theta_r = k_i \sin\theta_i + \frac{\partial \psi_{ms,r}}{\partial x} \tag{7}$$

$$k_t \sin\theta_t = k_i \sin\theta_i + \frac{\partial \psi_{ms,t}}{\partial x} \tag{8}$$

where $k_i = n_i \omega_i/c$ and

$$k_{\mathrm{s}} = \frac{n_s \omega_{\mathrm{s}}}{c} = \frac{n_s}{c}\left(\omega_{\mathrm{i}} - \frac{\partial \psi_{\mathrm{ms,s}}}{\partial t}\right), \quad \mathrm{s} = \{\mathrm{r,t}\}, \tag{9}$$

with $n_i (= n_r)$ and $n_t$ being the refractive indices of the incident and transmissive media, respectively. The forms of equations (7) and (8) which calculate angles of reflection and refraction are very similar to the equations introduced by space-gradient metasurfaces, but there is an important difference illustrated by formula (9). This equation implies that the amplitude of the wavenumbers is changing with the change of frequency (resulting from time variations of phase).

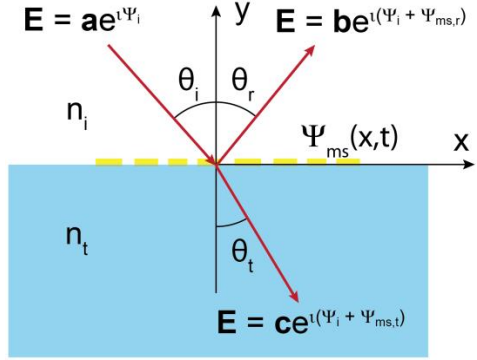

Fig. 1. Schematic of a light beam incident on a space-time gradient metasurface with angle of incidence $\theta_i$, reflected beam with angle of reflection $\theta_r$, and transmitted (refracted) beam with angle of transmission (refraction) $\theta_t$.

The above equations clearly indicate that the space-gradient phase-shift introduces an abrupt change to the photonic momentum with a value of $\Delta p_x = \hbar \Delta k_x = \hbar\, \partial \psi_{ms}/\partial x$, and similarly, a time-gradient phase-shift causes a photonic energy change of $\Delta E = \hbar \Delta \omega = -\hbar\, \partial \psi_{ms}/\partial t$. This Doppler-like shifting of the wavelength in time-varying metasurfaces enables alternative approaches that could be advantageous over mechanical movement. They can also be integrated with mechanical systems to modify or compensate the Doppler effect. Interesting and useful developments can be also related to cavity optomechanics[32], where time-varying metasurfaces could be utilized to expand the control over the inelastic photonic interactions with vibrating mirrors. Here the exchange of energy with photons could be used for laser cooling[33] or heating. Figure (2) illustrates a basic paradigm for the universal Snell's law with space-gradient and time-gradient metasurfaces. For the sake of simplicity and clarity, we discuss reflection from gradient metasurfaces in free space. A similar analysis can be extended to transmittance and for arbitrary media. Figure 2(a) demonstrates the photonic interaction with a space-gradient metasurface, where a discontinuity of tangential momentum (or wavenumber) is added to the reflected photons. Conservation of energy requires the amplitude of the total momentum to remain on the same isofrequency curve $\sqrt{k_x^2 + k_y^2} = \omega/c$ (or iso-energy curve $\sqrt{p_x^2 + p_y^2} = E/c$). The tangential and total momenta together define the new angle of reflection (solid line) which is different than the reflection angle with no metasurface (dashed line). Figure 2(b) demonstrates another degree of freedom to control the reflection of the beam using a time-gradient metasurface which, according to equation (5), introduces a change to the iso-frequency curve of the reflected photons. If there is no space-gradient phase shift introduced, then the tangential momentum would not change; however, the variation in frequency results in a change in the normal component of the momentum (solid line) compared to the case when the photon energy is conserved (dashed line). Figures 2(c-d) compare two examples of reflection tests upon time-reversal and demonstrate that unlike space-gradient metasurfaces, time-varying metasurfaces can provide non-reciprocal reflectance. Figure 2(c) illustrates that the variations induced in the tangential momentum by the space-gradient metasurface for the forward and reverse directions compensate each other. Thus, if a space-gradient metasurface adds a tangential momentum to the forward beam (Fig 2(a)), it subtracts the same amount in the reverse beam (Fig 2(c)), restoring the original incident direction. This is not the case for time-varying metasurfaces where the changes in the isofrequency curve are additive for both directions and do not negate each other. If the amplitude of the wavenumber is increased for the forward beam (Fig 2(b)), it is further increased to a higher value for the reversed beam (Fig 2(d)), leading to a deviation of the time-reversed reflected beam from the incident beam.

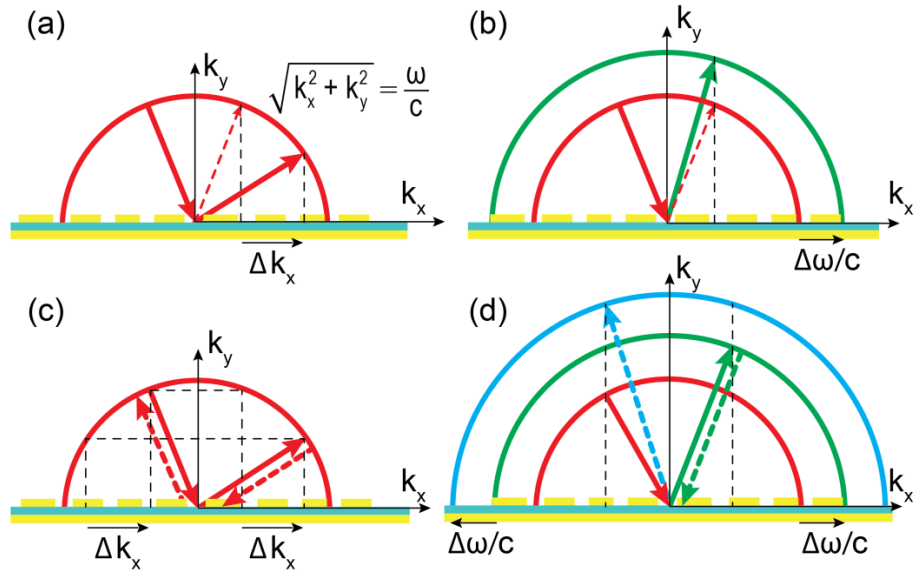

Fig.2. (a) Light reflected from a space-gradient metasurface inducing tangential momentum discontinuity. The dashed red line represents the reflected beam without the metasurface effect. (b) Light reflected from a time-gradient metasurface inducing energy discontinuity and changing the isofrequency curve. The dashed red line represents the reflected beam without the metasurface effect. (c) Time-reversal test of a space-gradient metasurface. Red dashed line denotes reciprocal propagation of light. (d) Performance of a time-gradient metasurface in time-reversal with the dashed green and blue lines denoting the nonreciprocal traces of the incident and reflected beams respectively.

Now, we quantify the amount of non-reciprocal deviation in the propagation direction for the case illustrated in Fig 2(d). For simplicity, we assume that there is no space-varying phase-shift ($\partial \psi_{ms}/\partial x = 0$), and that there is a linear variation of $\psi_{ms}$ with respect to time with a derivative value of $\Delta \omega = - \partial \psi_{ms}/\partial t$. This can be obtained by introducing a periodic phase shift that changes linearly from $\pi$ to $-\pi$ during a period $T = 2\pi/\Delta \omega$. Let the angles of incidence and reflection to this metasurface be $\theta_1$ and $\theta_2$ as shown in Fig 3(a). By applying equations (5) and (9), we derive that if the frequency and the wavenumber of the incident beam are $\omega$ and $k = \omega/c$, respectively, then the frequency and the wavenumber of the reflected beam are $\omega + \Delta\omega$ and $k + \Delta k = (\omega + \Delta\omega)/c$. Then, from equation (7), we obtain:

$$k \sin\theta_1 = (k + \Delta k)\sin\theta_2 \tag{10}$$

Using the same analysis for the time-reversal case shown in Fig 3(b) we find:

$$(k + \Delta k)\sin\theta_2 = (k + 2\Delta k)\sin\theta_3 \tag{11}$$

From equations (10) and (11) it follows that

$$\sin\theta_3 = \frac{\sin\theta_1}{1 + \frac{2\Delta k}{k}} = \frac{\sin\theta_1}{1 + \frac{2\Delta\omega}{\omega}} \tag{12}$$

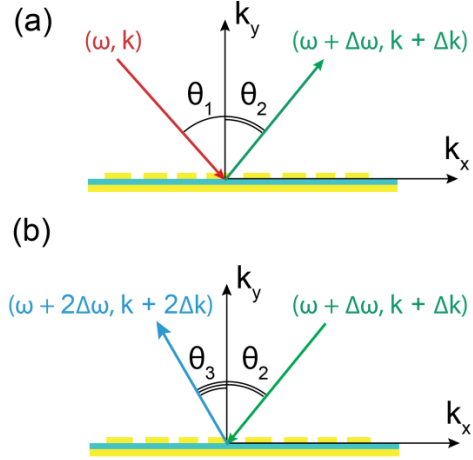

Fig. 3. (a) Schematics of reflection angle from a time-gradient metasurface. (b) reflection angle in time-reversal.

From equation (12) it can be concluded that the back reflected beam in time-reversal is not propagating in the same direction as the incident beam. Consequently, this type of metasurface can be used as a unidirectional isolator with two ports 1 and 2 placed along the directions of the incident beam at $\theta_1$ and the reflected beam at $\theta_2$, respectively. Light is allowed to propagate in the forward direction from port 1 to port 2, while light backscattered from port 2 is redirected at an angle away from port 1, enabling scattering parameters of $S_{21} > 0$ and $S_{12} \approx 0$, which is required for optical isolation[31]. Equation (12) demonstrates that the separation between $\theta_1$ and $\theta_3$ is proportional to the ratio $\Delta\omega/\omega$ which is very easy to control while operating in the radio frequency and possibly terahertz. For operation in the infrared and visible bands, considerable ratios of $\Delta\omega/\omega$ can be obtained when, for example, time-varying metasurfaces are modulated optically. We note that non-reciprocity attributed to the difference in photonic energy levels (frequencies) between the incident and back-scattered beams can be completely decoupled using trivial optical filtering with a high-quality-factor optical cavity. In this case even a small change in the frequency would provide an observable effect. Figures 4(a) and 4(b) depict the schematic of an optical isolator based on a metasurface with a frequency shift of $\Delta\omega = -\partial\psi_{ms}/\partial t$ and two optical resonators with center frequencies of $\omega$ and $\omega + \Delta\omega$. Figure 4(a) shows the allowed forward propagation for an incident beam of frequency $\omega$ and the reflected beam of frequency $\omega + \Delta\omega$, where both beams pass through the optical resonators. Figure 4(b) presents the backward propagation of the time-reversed $\omega + \Delta\omega$ beam, which is reflected at a shifted frequency of $\omega + 2\Delta\omega$ and hence, the reversed beam is blocked by the resonator.

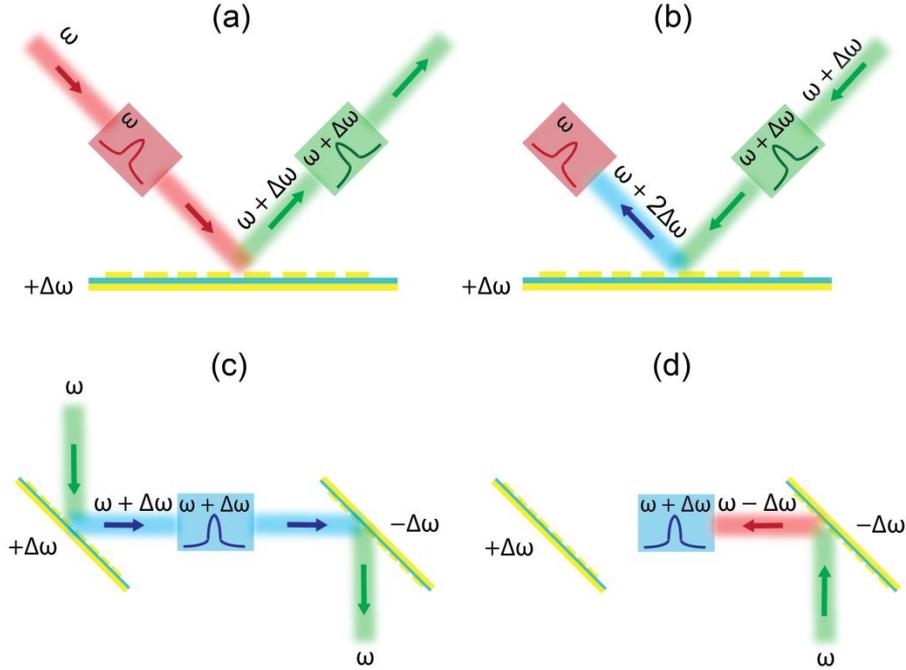

Fig. 4. (a,b) Schematics of an optical isolator with uni-directional light flow using a time-varying metasurface and two high quality resonators. (c,d) An optical isolator with same input/output frequency using two time-varying metasurfaces and a high quality resonator.

To build an isolator with the same input and output frequencies, one can use two metasurfaces with the same magnitude but opposite in sign phase-shifts. Thus, one achieves frequency shifts of $+\Delta\omega$ and $-\Delta\omega$, respectively, which compensate and restore the same frequency in the output. Inserting a resonator tuned at $\omega + \Delta\omega$ in the path of light between the two metasurfaces, would allow forward propagation of light (Figure 4(c)) and block its backward propagation (Figure 4(d)).

The examples proposed in Fig. 4 demonstrate straightforward approaches to designing compact, magnetic-free optical isolators with time-varying metasurfaces. Generally, these isolators would require either the utilization of non-linear or time-varying materials[31]. Due to some limitations on nonlinear isolators[34] such as intensity dependent operation, there is interest in developing isolators based on time-varying structures[35-38]. We believe that ultra-thin metasurfaces with dynamically induced phase-shift can be advantageous over bulk time-varying structures.

In conclusion, time-varying metasurfaces bring a new degree of freedom in controlling light with flat photonic devices that further broadens the scope of applications for metasurfaces. While space-gradient metasurfaces enable the control of the tangential momentum, space-time-gradient metasurfaces are capable of controlling both the photonic energy (normal momentum) and the tangential momentum. As a result, time-gradient metasurfaces go beyond the physical limitations imposed on space-gradient metasurfaces like reciprocity and 'elasticity' of the reflected and refracted photons. We illustrated that time-gradient metasurfaces can break Lorentz reciprocity, enabling the development of optical isolators, and derived a universal non-reciprocal form of Snell's law. Photons interacting with time-varying metasurface exhibit Doppler-like wavelength shift causing exchange of energy. This is similar to the effect induced by vibrating mirrors in opto-mechanical cavities, where energy exchange with photons is used, for example, for laser cooling. Additionally, time-varying metasurfaces can be integrated with time-reversal mirrors[39-41] and used for subwavelength focusing[42] enabling the mapping of backward-signals or far-field images. For radio-frequency and microwave applications, the temporal modulation of metasurfaces can be achieved for example by using an array of varactor-based phase-shift elements. Optical implementations are possible by using, for example, electroptic and photoacoustic effects as well as other available modulation techniques.

**Acknowledgement**

This work was partially supported by AFOSR grant 123885-5079396, ARO grant W911NF-13-1-0226, and NSF grant DMR-1120923.